\documentclass[aps,prxquantum,reprint,superscriptaddress,amsmath,amssymb,pra,showkeys,longbibliography]{revtex4-2}

\usepackage{graphicx}%
\usepackage{dcolumn}%
\usepackage{bm}%
\usepackage[colorlinks=true,linkcolor=blue, citecolor=blue, urlcolor=blue]{hyperref}%
\usepackage{braket}
\usepackage{appendix}
\usepackage{natbib}
\usepackage{physics}
\usepackage{float}
\usepackage{booktabs}
\usepackage[T1]{fontenc}
\usepackage{dsfont}
\usepackage{adjustbox}
\usepackage{multirow}
\usepackage{gensymb}
\usepackage{array}
\usepackage{siunitx}
\usepackage{xcolor}
\usepackage{xr}
\usepackage[sort&compress]{cleveref}
\usepackage[caption=false]{subfig}

\newcolumntype{L}{>{$}l<{$}}
\newcolumntype{C}{>{$}c<{$}}
\newcolumntype{R}{>{$}r<{$}}

\usepackage[hang,flushmargin,symbol*]{footmisc}
\DefineFNsymbolsTM{otherfnsymbols}{
	\textdagger    \dagger
}

\colorlet{darkgreen}{green!40!black}

\begin{document}

	\title{Switchable heavy-hole/light-hole spin qubit}
	\author{Zoltán György}
    \email{zoltan.gyoergy@unibas.ch}
\affiliation{Department of Physics, University of Basel, Klingelbergstrasse 82, CH-4056 Basel, Switzerland}

	\author{Dmitry Miserev}
\affiliation{Department of Physics, University of Basel, Klingelbergstrasse 82, CH-4056 Basel, Switzerland}

    \author{Jelena Klinovaja}
\affiliation{Department of Physics, University of Basel, Klingelbergstrasse 82, CH-4056 Basel, Switzerland}

    \author{Daniel Loss} 
\affiliation{Physics Department, King Fahd University of Petroleum and Minerals, 31261, Dhahran, Saudi Arabia}
\affiliation{Quantum Center, KFUPM, Dhahran, Saudi Arabia}
\affiliation{RDIA Chair in Quantum Computing}

	\date{\today}            
	
	\begin{abstract}
Compressively strained Ge quantum wells in planar SiGe/Ge heterostructures are the state-of-the-art platform for hole spin qubits. While they exhibit robust coherence times, they possess weak intrinsic spin-orbit interaction (SOI) due to the heavy-hole (HH) character of the wavefunction. Recently, light-hole (LH) qubits were proposed in GeSn/Ge heterostructures, offering strong, intrinsic, linear-in-momentum SOI. In this work, we propose a switchable HH-LH spin qubit in a bilayer Ge heterostructure with SiGeSn barriers, combining the advantages of HH and LH devices. The character of the qubit can be changed by shuttling from an LH well to an HH well, which also enables fast, hopping-based single-qubit rotations. Additionally, we observe an HH-LH resonance introduced by the in-plane confinement, resulting in $g$-factor peaks and first-order charge noise sweet spots. Our calculations reveal a sweet spot with Rabi frequencies on the order of 100 MHz, comparable to the LH regime, but with a more than tenfold increase in coherence time, on the order of 100 $\mu$s. %

    \end{abstract}
    \maketitle

\section{\label{sec:intro}Introduction}
The pursuit of fault-tolerant quantum computation relies heavily on scalable physical platforms. In this regard, semiconductor-based spin qubits \cite{PhysRevA.57.120, Fang_2023, RevModPhys.95.025003,mcintyre2026theory} have emerged %
 as highly competitive candidates, largely driven by their potential for high-density integration \cite{philips2022universal, hendrickx2021four,borsoi2024shared,zhang2025universal,wang2024operating, john2024two,george202412, lim20242x2} and the ability to leverage established CMOS fabrication techniques \cite{maurand2016cmos,zwerver2022qubits,steinacker2025industry,george202412}. Furthermore, these platforms demonstrate high-fidelity single- and two-qubit operations \cite{veldhorst2014addressable,yoneda2018quantum,lawrie2023simultaneous,mills2022high,huang2024high, wu2025simultaneous,xue2022quantum,mills2022two,noiri2022fast,huang2024high}. Crucially, spin qubits also support the long-range connectivity required for advanced error-correction protocols, such as qLDPC codes \cite{andersen2025small}. The required long-range entanglement can be mediated either through coherent shuttling \cite{yoneda2021coherent,kunne2024spinbus,bosco2024high,van2024coherent,de2025high,ademi2025distributing} or via the integration of superconducting microwave resonators \cite{mi2018coherent,zwerver2022qubits,eggli2025coupling,noirot2025coherence}.

\begin{figure}[b]
\centering
\includegraphics[width=\columnwidth]{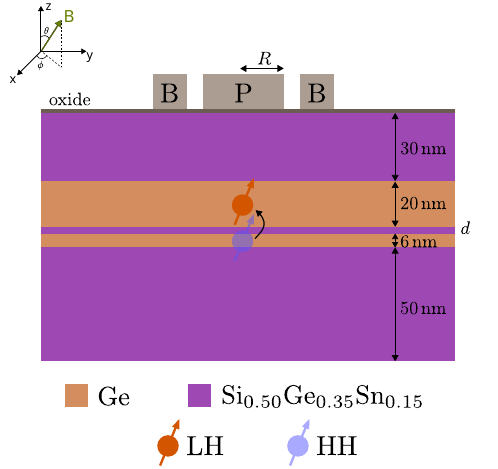}

\caption {\label{fig:device}Heterostructure for switchable heavy-hole/light-hole spin qubit. The SiGeSn barriers are assumed to be unstrained, while the Ge quantum wells are slightly tensile-strained. A single hole is confined in the lower layer, where it has heavy-hole character.
 After shuttling it to the upper Ge layer, it becomes a light-hole qubit. The two barrier gates (B) provide the in-plane confinement, while the plunger gate (P) is used for shuttling.}
\end{figure}

At present, the most advanced spin qubit architectures are realized using silicon-based electrons \cite{philips2022universal,fernandez2026running,undseth2026weight} or nuclear spins \cite{edlbauer202511}, alongside holes confined in germanium planar heterostructures \cite{john2024two,dijkema2026simultaneous}. This latter platform---germanium hole qubits---offers a unique set of physical advantages. Because holes possess a lighter effective mass than electrons, the lithographic constraints for defining dense quantum dot arrays are significantly relaxed. Additionally, their $p$-orbital symmetry suppresses hyperfine decoherence from host nuclei, while an intrinsic spin-orbit interaction (SOI) facilitates rapid, all-electrical qubit driving via electric-dipole spin resonance (EDSR) without the need for micromagnets \cite{bulaev2007electric,terrazos2021theory,wang2022ultrafast,martinez2022hole,abadillo2023hole}. Finally, the hole $g$-tensor \cite{crippa2018electrical,venitucci2018electrical} is highly sensitive to local electric fields \cite{liles2021electrical,bassi2024optimal,mauro2025hole,seidler2025spatial, sommer2026disentangling}. By exploiting this site-to-site $g$-factor variability, architectures can implement baseband-driven single-qubit control using hopping spins \cite{wang2024operating}. To date, experimental demonstrations of hole spin qubit arrays have relied on compressively strained Ge quantum wells with SiGe barriers \cite{hendrickx2021four,john2024two,zhang2025universal,dijkema2026simultaneous}. The compressive strain lifts the heavy-hole/light-hole degeneracy, leading to a heavy-hole (HH) ground state and a large heavy-hole/light-hole splitting \cite{terrazos2021theory,wang2024modeling}. These heavy-hole qubits possess a weak intrinsic cubic Rashba SOI, while the dominant mechanism for fast EDSR is linear SOI originating from shear strain gradients \cite{abadillo2023hole} or interface effects \cite{rodriguez2023linear,sarkar2025effect}.

Although strain and heterostructure engineering have been investigated for SiGe/Ge quantum wells \cite{nigro2025strain,mauro2025strain,del2026tailoring,shojaei2026g}, research in this domain remains in its infancy. As novel material compositions and new heterostructure designs become experimentally viable, it is crucial to explore the unique physical properties and new device architectures they can enable. Notable examples include tensile-strained GeSn/Ge quantum wells \cite{assali2022light}, which are anticipated to host light-hole (LH) spin qubits \cite{del2023light,de2024strong} with large in-plane $g$-factors and strong linear Rashba SOI, resulting in fast single-qubit operations. Another example is the experimental demonstration of unstrained SiGe/Ge quantum wells \cite{costa2025buried}, leading to a heavy-hole ground state with a small heavy-hole/light-hole splitting \cite{mauro2025hole}. Very recently, tensile-strained SiGe quantum wells lattice-matched to a Ge substrate were proposed to host LH spin qubits without Sn incorporation~\cite{valvo2026lightholeARXIV}. Finally, bilayer Ge heterostructures have been experimentally demonstrated, hosting up to eight qubits in 2D or 3D arrangements \cite{ivlev2024coupled,tidjani2025three}.

Building upon these advancements in heterostructure and strain engineering, we theoretically propose a bilayer heterostructure design to combine the advantages of heavy-hole and light-hole qubits. A complementary route to an electrically switchable HH-LH qubit, stacking a tensile-strained SiGe well on top of an unstrained Ge channel, was proposed in concurrent work~\cite{valvo2026lightholeARXIV}. A single hole in such a bilayer device exhibits a hybrid character: depending on its spatial localization between the two wells, it has predominantly heavy-hole, light-hole, or strongly mixed character.
 We show that fast single-qubit operations can be realized using EDSR in the light-hole regime, while the heavy-hole state can be used for coherent idling. Additionally, fast single-qubit gates can be achieved using hopping, owing to the large $g$-factor differences between the two wells. Finally, we identify charge-noise sweet spots in the highly mixed heavy-hole/light-hole regime, simultaneously yielding long coherence and fast operation times. 

This manuscript is structured as follows. In Sec.~\ref{sec:model}, we introduce the heterostructure design, our model, and assumptions. In Sec.~\ref{sec:g-factor}, we study the $g$-tensor anisotropy and the heavy-hole/light-hole mixing of the device. In Sec.~\ref{sec:hopping}, we explore hopping-based single-qubit operations, while in Sec.~\ref{sec:EDSR} we present spin manipulation based on EDSR. In Sec.~\ref{sec:noise}, we introduce a simple model for the charge noise and study the qubit coherence, identifying charge-noise sweet spots. Finally, we discuss our results and conclude in Sec.~\ref{sec:discussion}.

\section{Heterostructure and theoretical model}\label{sec:model}
The proposed epitaxially grown heterostructure is schematically represented in Fig.~\ref{fig:device}. A single hole can be confined in either of the two Ge quantum wells, which are separated by an unstrained $\mathrm{Si}_{0.50}\mathrm{Ge}_{0.35}\mathrm{Sn}_{0.15}$ barrier. This specific ternary alloy composition induces a small tensile strain of 0.25\% in the Ge layers. While this tensile strain typically favors a light-hole ground state with a small heavy-hole/light-hole splitting, the ultimate character of the hole is also governed by quantum confinement due to the different effective masses of the two states. Because strong confinement energetically favors heavy holes, this effect can overcome the small strain potential. Consequently, the narrow 6 nm quantum well yields a heavy-hole ground state, whereas the wider 20 nm well can host a light hole. The character of the state can be changed by shuttling the hole using the plunger gate (P), while the in-plane confinement is ensured by the two barrier gates (B). 

\begin{figure}[H]
\centering
\includegraphics[width=\columnwidth]{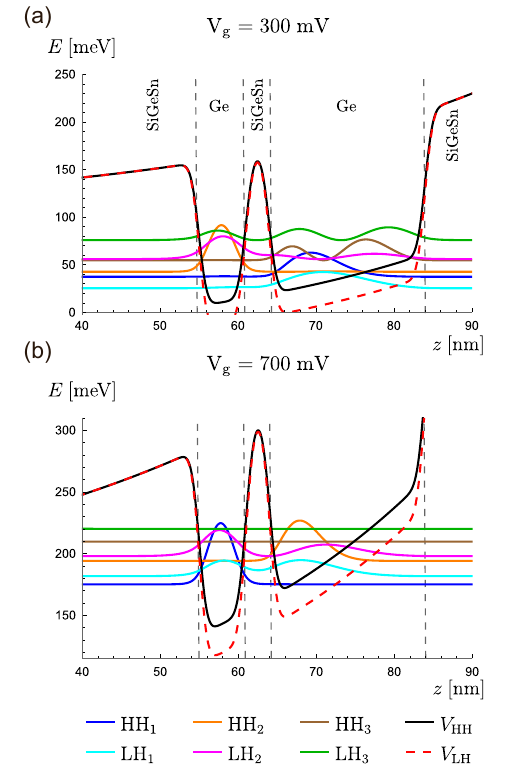}

\caption{\label{fig:QW}Quantum well confinement and lowest eigenstates. The quantum well confinement profile is shown for heavy holes (black curve) and light holes (red dashed line), together with the lowest three heavy-hole and light-hole eigenstates from Eq.~\eqref{eq:QWsolution}. By changing the gate voltage $V_g$, we can tune the character of the ground state from (a) light hole ($\mathrm{LH}_1$) to (b) heavy hole ($\mathrm{HH_1}$). }
\end{figure}

SiGeSn alloys have emerged as a highly compelling class of materials, due to their applications in group-IV laser physics \cite{sun2010design,ghetmiri2017study,stange2018gesn} and in tunnel field-effect transistors \cite{wirths2013band}. Here, we demonstrate the potential of SiGeSn alloys as a versatile and promising platform for heterostructure and strain engineering of hole spin qubits. For our proposed design, achieving the required alloy composition of 50\% Si and 15\% Sn is experimentally demanding due to the large lattice mismatch between Si and Sn. However, SiGeSn heterostructures with concentrations as high as 48\% Si and 12\% Sn have been successfully demonstrated \cite{shibayama2025emergence,zhang42epitaxial}, indicating the practical feasibility of our architecture.

The reason for using the ternary alloy SiGeSn for the barriers is twofold. First, it enables precise engineering of the lattice constant, thereby controlling the strain in the Ge quantum wells. Second, a high Si concentration is required to achieve a large valence band offset relative to Ge, allowing the confinement of both heavy and light holes. By contrast, a GeSn barrier could induce the target tensile strain but would fail to confine the heavy-hole states \cite{del2025fully}.

We model the device from Fig.~\ref{fig:device} using the 6-band Luttinger-Kohn-Bir-Pikus (LKBP) Hamiltonian \cite{luttinger1956quantum,bir1974symmetry,winkler2001spin}: 
\begin{equation}\label{eq:H3D}
    H_\mathrm{3D}=H_\mathrm{LKBP}(k_x,k_y,k_z)+V_\mathrm{qw}(z)+V_E(z)+V(x,y),
\end{equation}
where $H_\mathrm{LKBP}(k_x,k_y,k_z)$ is the 6-band LKBP Hamiltonian assuming only biaxial strain due to the lattice mismatch between Ge and the relaxed SiGeSn. $V_\mathrm{qw}(z)$ describes the quantum well confinement via the valence band offsets, $V_E(z)$ is the potential created by the plunger gate (P, see Fig.~\ref{fig:device}), and $V(x,y)$ is the in-plane confinement created by the two barrier gates. We choose our coordinate system such that the $x$, $y$, and $z$ axes align with the [100], [010] and [001] crystal axes, respectively. We work in the hole representation, with positive hole effective masses.

Many theoretical studies assume a simple analytical potential for $V_E(z)$ and a harmonic confinement for $V(x,y)$ \cite{wang2024modeling,sarkar2025effect,terrazos2021theory,del2023light,del2024light,del2025fully}, while others solve Poisson's equation with more realistic gate layouts, focusing on conventional SiGe/Ge heterostructures \cite{rodriguez2023linear,mauro2025strain,martinez2022hole,abadillo2023hole}. Owing to the novelty of our proposed material stack, we employ the former approach. We assume an anisotropic harmonic in-plane confinement potential: 
\begin{equation}\label{eq:in-plane}
    V(x,y)=\frac{m_c}{2}(\omega_x^2x^2+\omega_y^2y^2),
\end{equation}
where $m_c$ is a phenomenological mass parameter describing the in-plane confinement potential (not to be confused with the hole effective mass), and $\omega_x$ and $\omega_y$ describe the confinement strengths. Throughout the calculations, we set $m_c=0.5m_0$, where $m_0$ is the free electron mass. For the confinement created by the plunger gate, we assume the following
\begin{equation}\label{eq:Vz}
    V_E(z)=\frac{2e V_g}{\pi}\arctan{\left(\frac{R}{z}\right)},
\end{equation}
which is the potential at distance $z$ from the center of a metallic (equipotential) disk of radius $R$ (see Fig.~\ref{fig:device}) held at voltage $V_g$
 \cite{friedberg1993electrostatics}. Throughout the calculations, we set $R=35$ nm. This analytical function neglects the different permittivities of Ge and SiGeSn, but decays to zero at large distances, in contrast to the constant electric field approximation employed in many works \cite{wang2024modeling,sarkar2025effect,terrazos2021theory,del2023light,del2024light,del2025fully}. Having a potential that decays to zero is important, considering the large heterostructure in Fig.~\ref{fig:device}, where we include only 50 nm of the bottom SiGeSn barrier in the calculations.

We calculate the material parameters of $\mathrm{Si}_{0.50}\mathrm{Ge}_{0.35}\mathrm{Sn}_{0.15}$ by linearly interpolating the respective concentrations of Si, Ge, and Sn. However, linear interpolation for the Luttinger parameters $\gamma_1$, $\gamma_2$ and $\gamma_3$ is known to be inaccurate \cite{lu2012electronic, del2025fully}. As the correct Luttinger parameters are unknown for SiGeSn alloys with varying compositions, we use the Luttinger parameters of Ge for the whole heterostructure. Because the hole wavefunction is confined within the Ge quantum well, approximating the barrier's Luttinger parameters with those of Ge is reasonable. More details about the 6-band LKBP Hamiltonian and the material parameters used in this work can be found in Appendix~\ref{App:parameters}.

Owing to the thin Ge quantum well and SiGeSn barrier of the heterostructure in Fig.~\ref{fig:device}, we assume continuous compositional profiles \cite{sammak2019shallow} instead of perfectly sharp material interfaces
\begin{equation}
    c_i(z)=\frac{c_{i,L}+c_{i,R}}{2}+\frac{c_{i,R}-c_{i,L}}{2}\mathrm{erf}\left(\frac{z-z_0}{\sqrt{2}\sigma}\right),
\end{equation}
where $i$ is the material, $i\in$\{Si, Ge, Sn\}, $z_0$ is the position of the interface, $c_{i,L(R)}$ is the concentration to the left (right) of the interface, and $\sigma$ describes the width of the interface. For all the calculations, we set $\sigma=0.8$ nm.

The full, 3D Hamiltonian from Eq.~\eqref{eq:H3D} contains separable out-of-plane and in-plane confinement potentials. Therefore, we can eliminate the $z$-dependence and construct the Hamiltonian of the two-dimensional hole gas (2DHG) by finding the following eigenstates 
\begin{equation}\label{eq:QWsolution}
    H_\mathrm{QW}(z,k_z)\phi_i^{(\alpha)}(z)=E_i^{(\alpha)}\phi_i^{(\alpha)}(z),
\end{equation}
where $H_\mathrm{QW}(z,k_z)$ is 
\begin{equation}
\begin{aligned}
    H_\mathrm{QW}(z,k_z)&=H_\mathrm{LKBP}(k_x=0,k_y=0,k_z)\\&+V_\mathrm{qw}(z)+V_E(z),
\end{aligned}
\end{equation}
and $\alpha\in\{\mathrm{H},\mathrm{L}\}$. The $\phi_i^{(H)}(z)$ functions are heavy-hole envelope functions, while $\phi_i^{(L)}(z)$ are mixed light-hole and split-off hole envelope functions. However, the low-energy mixed eigenfunctions have predominantly light-hole character, so we simply refer to them as light-hole eigenfunctions. We can construct the 2DHG Hamiltonian $H_\mathrm{2D}(k_x,k_y)$ by expanding $H_\mathrm{LKBP}(k_x,k_y,k_z)+V_\mathrm{qw}(z)+V_E(z)$ on the envelope functions $\phi_i^{(\alpha)}(z)$ \cite{del2025fully}.

Figure~\ref{fig:QW} shows the quantum well confinement and the functions $\phi_i^{(\alpha)}$ with the lowest energies, for two different plunger gate voltages $V_g$. Due to the small tensile strain, the heavy holes and light holes feel different confinement potentials, $V_\mathrm{HH}$ and $V_\mathrm{LH}$, respectively. At $V_g=300$ mV, the ground state has light hole character, while at $V_g=700$ mV, the ground state is a heavy hole state. 

We note that the material stack shown in Fig.~\ref{fig:device} requires positive plunger gate voltages to access the heavy-hole regime, since the heavy-hole well is situated further from the gates than the light-hole well. This layout offers a distinct advantage: the idling heavy-hole spin is spatially further separated from the oxide interface and its associated two-level fluctuators (TLFs), thereby prolonging the coherence time. Conversely, confining a single hole becomes more complex, as the plunger gate voltage must be positive while the barrier gate voltages remain negative to ensure in-plane confinement. An alternative arrangement is possible, where the heavy-hole well is placed closer to the oxide; this allows for standard negative plunger voltages but compromises the heavy-hole coherence. Throughout the main text, we focus on the former architecture, while in Appendix~\ref{App:Alternative} we discuss the latter.

\section{\texorpdfstring{$g$}{g}-factor anisotropy and heavy-hole/light-hole resonance}\label{sec:g-factor}
Now, we turn to the properties of the quantum dot. We numerically solve the eigenvalue problem from Eq.~\eqref{eq:QWsolution} using finite-difference discretization. We expand the full Hamiltonian from Eq.~\eqref{eq:H3D} on the lowest 150 heavy-hole and lowest 150 light-hole eigenstates of Eq.~\eqref{eq:QWsolution}, sufficient for convergence to within 1\%, and obtain the 2DHG Hamiltonian $H_\mathrm{2D}(k_x,k_y)$ \cite{del2025fully}. The numerical solution of the quantum well confinement spans the region shown in Fig.~\ref{fig:device}, with a grid resolution of 0.033 nm. The quantum dot Hamiltonian can be written as 
\begin{equation}
    H_\mathrm{dot}=H_\mathrm{2D}(k_x,k_y)+\frac{m_c}{2}(\omega_x^2x^2+\omega_y^2y^2). 
\end{equation}
We solve this Hamiltonian in momentum space, using finite-difference discretization. We discretize $(k_x,k_y)$ in $100 \times 100$ points ($80\times 80$ for calculations with finite magnetic field), with $\lvert k_{x(y)} \rvert<0.4$ to $0.6$ $\mathrm{nm^{-1}}$, depending on in-plane confinement strength. To extract the $g$-tensor, we first solve the Hamiltonian in the absence of a magnetic field to obtain the Kramers-degenerate qubit states $\ket{\Uparrow}$ and $\ket{\Downarrow}$. Then, using the linear-in-$B$ terms of the full Hamiltonian from Eq.~\eqref{eq:H3D} (including both direct Zeeman terms and orbital contributions), we construct the $g$-tensor, as described in Refs. \cite{crippa2018electrical,venitucci2018electrical}. Since our model has no inhomogeneous strain, the $g$-tensor is diagonal in the $(x,y,z)$ frame; the effective $g$-factors $g_x$, $g_y$, and $g_z$ plotted below are its principal values, $g_i=\lvert g\,\hat{e}_i\rvert$, i.e., the Larmor frequency is $\mu_B g_i B/h$ for $\mathbf{B}=B\hat{e}_i$. The $g$-tensor is independent of $B$; the quantization-axis misalignment angle of Sec.~\ref{sec:hopping} depends on the field direction but not on its magnitude. Wherever the field magnitude matters (Secs.~\ref{sec:EDSR} and \ref{sec:noise}) we use $B=0.1$ T. 

\begin{figure}[htp!]
\centering
\includegraphics[width=\columnwidth]{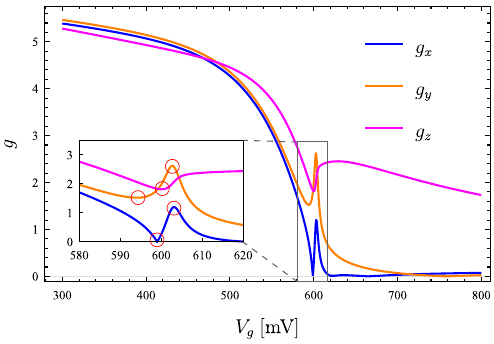}

\caption {\label{fig:g1}Effective qubit $g$-factors as a function of plunger gate voltage $V_g$. The magnetic field is along the $x$ (blue line), $y$ (orange line), or $z$ (magenta line) direction. The qubit transitions from the light-hole regime to the heavy-hole regime. In the heavily mixed regime, around $V_g=602.7$ mV, the $g$-factors show nonmonotonic behavior. The red circles show first-order charge noise sweet spots. The confinement frequencies are $\hbar\omega_x=2$ meV, $\hbar\omega_y=2.6$ meV, and the barrier width is $d=3$ nm (see Fig.~\ref{fig:device}).}
\end{figure}

\begin{figure}[htp!]
\centering
\includegraphics[width=\columnwidth]{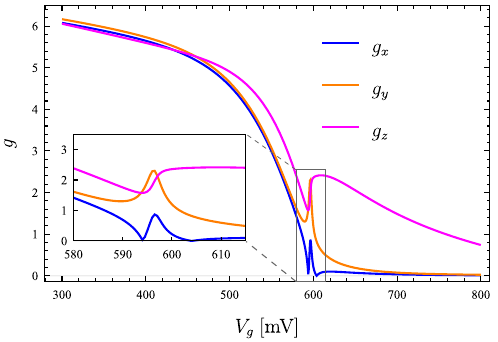}

\caption {\label{fig:g3}Effective $g$-factors as a function of plunger gate voltage $V_g$. The magnetic field is along the $x$ (blue line), $y$ (orange line), or $z$ (magenta line) direction. The confinement frequencies are $\hbar\omega_x=1.5$ meV, $\hbar\omega_y=1.95$ meV, and the barrier width is $d=3$ nm (see Fig.~\ref{fig:device}).}
\end{figure}

The effective $g$-factors are plotted in Fig.~\ref{fig:g1} as a function of plunger gate voltage $V_g$, when the magnetic field is along the $x$, $y$, or $z$ direction. The anisotropic in-plane confinement frequencies are $\hbar\omega_x=2$ meV and $\hbar\omega_y=2.6$ meV, and the size of the barrier separating the two quantum wells is $d=3$ nm. By tuning the qubit from the light-hole ($V_g=300$ mV) to the heavy-hole regime ($V_g=800$ mV), the initially large in-plane $g$-factors are effectively quenched. The striking observation, however, is that this transition is nonmonotonic: the in-plane $g$-factors peak at $V_g=602.7$ mV. In the heavy-hole regime, the in-plane $g$-factors are close to zero; moreover, $g_x$ vanishes exactly at $V_g\approx 600$ mV in the mixed regime (inset of Fig.~\ref{fig:g1}). %
 The out-of-plane $g$-factor $g_z$ has a value around 2, which is small compared to standard heavy-hole qubits. %
 This is the result of the strong confinement in the narrow, 6 nm quantum well. Furthermore, because the HH-LH splitting remains small even at $V_g=800$~mV, $g_z$ is also sensitive to the strength of the in-plane confinement (cf.~Fig.~\ref{fig:g3}).

\begin{figure}[htp!]
\centering
\includegraphics[width=\columnwidth]{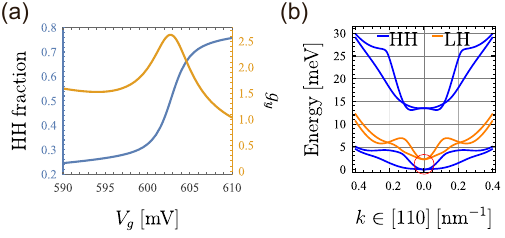}

\caption{\label{fig:Mixing}HH-LH resonance and $g$-factor peak. (a) The $g$-factor $g_y$ as a function of gate voltage (orange curve) and the heavy-hole fraction of the quantum dot ground state (blue curve). At the $g$-factor peak, the qubit has a fully hybridized HH-LH state, with 50\% HH fraction. The confinement frequencies are $\hbar\omega_x=2$ meV and $\hbar\omega_y=2.6$ meV, the barrier width is $d=3$ nm. (b) 2DHG spectrum at the gate voltage corresponding to the $g$-factor peak, $V_g=602.7$ mV. The blue curves are heavy-hole subbands, and the orange curves are light-hole subbands. The red circle shows the four spin-split subbands participating in the HH-LH resonance.}
\refstepcounter{subfigure}\label{fig:Mixinga}
\refstepcounter{subfigure}\label{fig:Mixingb}
\end{figure}
To shed light on the origin of the $g$-factor peaks in Fig.~\ref{fig:g1}, we plot the HH fraction (weight of $\ket{\frac{3}{2},\pm\frac{3}{2}}$ spinors) of the qubit eigenstates with a blue line in Fig.~\ref{fig:Mixinga}.
 The qubit transitions from the LH regime with an approximately 20\% HH fraction to the HH regime, with an approximately 80\% HH fraction within the voltage window shown. At the $g$-factor peak, the qubit has a fully hybridized HH-LH state, with 50\% HH fraction. In the LH (HH) regime, at $V_g=300$ (800) mV, the eigenstate has 9.3\% (92.0\%) HH fraction.

The observed $g$-factor peak and the maximally hybridized qubit state originate from an HH-LH resonance induced by the in-plane confinement. Figure~\ref{fig:Mixingb} shows the 2DHG subbands at the gate voltage corresponding to this peak ($V_g=602.7$ mV). The lowest two spin-split subbands exhibit heavy-hole character (blue lines), followed by two light-hole subbands (orange lines). Because heavy holes possess a smaller in-plane effective mass than light holes, the lateral confinement selectively elevates the HH subbands relative to the LH subbands. This relative shift drives the system into resonance, yielding a fully hybridized qubit state. 

Assuming parabolic HH and LH subbands with isotropic in-plane effective masses $m_\mathrm{HH}^\parallel=m_0/(\gamma_1+\gamma_2)$ and $m_\mathrm{LH}^\parallel=m_0/(\gamma_1-\gamma_2)$, the relative shift of the HH-LH splitting by the in-plane confinement is $\frac{\hbar(\omega_x+\omega_y)}{2}\big(\sqrt{m_c/m^{\parallel}_\mathrm{HH}}-\sqrt{m_c/m^{\parallel}_\mathrm{LH}}\big)=1.91$ meV. 
The energy gap between the relevant HH and LH subbands (highlighted by the red circle in Fig.~\ref{fig:Mixingb}) is approximately 2.2 meV, in good agreement with this estimate, confirming the resonance mechanism. %

\begin{figure}[htp!]
\centering
\includegraphics[width=\columnwidth]{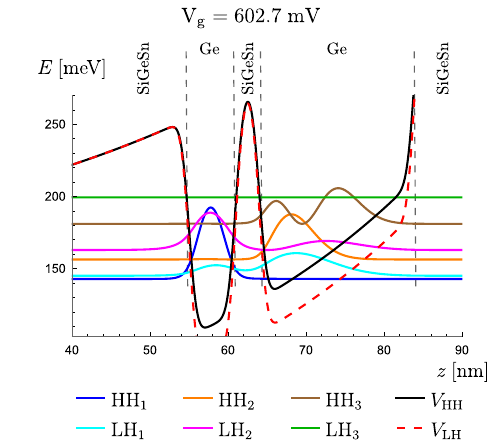}

\caption {\label{fig:QWRes}Quantum well confinement and eigenstates at the HH-LH resonance, at $V_g=602.7$ mV with a barrier width $d=3$ nm. The HH-LH splitting between the lowest HH (dark blue) and the lowest LH (light blue) state is small enough for the resonance to be induced by the in-plane confinement. There is also substantial overlap between these wavefunctions in the thin Ge well.}
\end{figure}

\begin{figure}[htp!]
\centering
\includegraphics[width=\columnwidth]{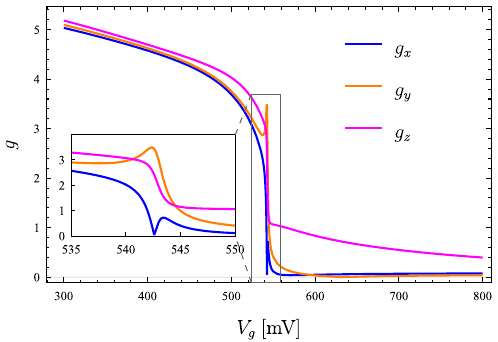}
\caption {\label{fig:g2}Effective $g$-factors as a function of plunger gate voltage $V_g$. The magnetic field is along the $x$ (blue line), $y$ (orange line), or $z$ (magenta line) direction. The confinement energies are $\hbar\omega_x=2$ meV, $\hbar\omega_y=2.6$ meV, and the barrier width is $d=4$ nm (see Fig.~\ref{fig:device}).}
\end{figure}

A necessary condition for the HH-LH resonance—beyond the resonant in-plane confinement—is a substantial overlap between the HH and LH quantum well eigenfunctions, $\phi_i^{(\alpha)}(z)$ (see Fig.~\ref{fig:QW}). A narrow barrier between the two quantum wells ($d=3$ nm, Fig.~\ref{fig:g1}) provides strong overlap, yielding a continuous transition from the LH to the HH regime and a pronounced $g$-factor peak. However, for wider barriers ($d=4$ nm, Fig.~\ref{fig:g2}), this overlap becomes exponentially suppressed, leading to a more abrupt transition. Figure~\ref{fig:QWRes} shows the quantum well wavefunctions $\phi_i^{(\alpha)}(z)$ at the HH-LH resonance for a barrier width of $d=3$ nm. There is a substantial overlap of the HH and LH wavefunctions in the thin Ge quantum well.

The magnitude of the $g$-factor peak is sensitive to the strength of the in-plane confinement, becoming more pronounced as the confinement tightens. Figure~\ref{fig:g3} illustrates the system in a weaker confinement regime ($\hbar\omega_x = 1.5$ meV, $\hbar\omega_y = 1.95$ meV). Compared to the stronger confinement scenario shown in Fig.~\ref{fig:g1} ($\hbar\omega_x = 2.0$ meV, $\hbar\omega_y = 2.6$ meV), the amplitudes of the $g$-factor peaks are visibly reduced.

\section{Hopping}\label{sec:hopping}
Hopping is an efficient way to implement single-qubit gates on hole spin qubits, relying solely on baseband pulses and site-dependent $g$-tensors \cite{wang2024operating,martinez2026disorder}. This principle---rotating a spin by displacing it into a region with a different $g$-tensor---goes back to~\cite{PhysRevA.57.120} and was recently extended to hopping-based classical reversible logic~\cite{loss2026reversible}. Here, we propose hopping between the two quantum wells of our bilayer heterostructure, rather than the conventional approach of hopping between neighboring quantum dots in a single layer \cite{wang2024operating,martinez2026disorder}. Vertical-shuttling single-qubit gates of this kind were recently proposed for electron spin qubits in Si/SiGe vertical double quantum dots~\cite{sarkar2026micromagnet}. Due to the significantly different $g$-tensors of heavy and light holes, hopping is expected to be highly efficient in our device. 

\begin{figure}[htp!]
\centering
\includegraphics[width=\columnwidth]{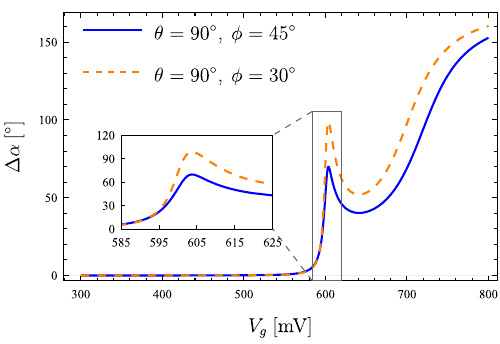}

\caption {\label{fig:alpha}Misalignment angle of the quantization axes as a function of gate voltage.
 The reference quantization axis is at $V_g=300$ mV. The angles are calculated using Eq.~\eqref{eq:misalignment}. The barrier width is $d=3$ nm, the confinement frequencies are $\hbar\omega_x=2$ meV and $\hbar\omega_y=2.6$ meV. The two curves show different in-plane directions of the magnetic field, see Fig.~\ref{fig:device}.}
\end{figure}

Hopping is achieved via repeated shuttling of the spin between the two layers, with the number of shuttles depending on the angle $\Delta\alpha$ between the two quantization axes. If the angle is larger than $22.5^\circ$, four shuttling steps are sufficient \cite{wang2024operating}. We calculate the angle $\Delta\alpha$ as follows

\begin{equation}\label{eq:misalignment}
\cos{(\Delta\alpha)}=\frac{(g_1\mathbf{B})\cdot(g_2\mathbf{B})}{\lvert g_1 \mathbf{B} \rvert \lvert g_2 \mathbf{B} \rvert},
\end{equation}
where $g_{1(2)}$ are the two $g$-tensors, and $g_{1(2)}\mathbf{B}$ gives the first (second) Larmor vector. The $g$-tensors, however, are defined only up to an arbitrary SU(2) rotation of the Kramers-degenerate qubit states $\ket{\Uparrow}$ and $\ket{\Downarrow}$. To resolve this ambiguity, we transform the eigenstates to align the $g$-tensors \cite{venitucci2018electrical,martinez2026disorder} at different gate voltages $V_g$ with the reference $g$-tensor at $V_g=300$ mV. 

The calculated angle between the quantization axes is shown in Fig.~\ref{fig:alpha}, for in-plane magnetic fields, with a barrier width $d=3$ nm and confinement frequencies $\hbar\omega_x=2$ meV and $\hbar\omega_y=2.6$ meV. The angle remains small before the HH-LH transition, at which point it abruptly increases. Because our model does not include inhomogeneous strains, the principal magnetic axes coincide with the symmetry axes of the confinement. Consequently, to achieve a large angle $\Delta \alpha$, the applied magnetic field cannot align with the $x$, $y$, or $z$ symmetry axes.

\begin{figure}[htp!]
\centering
\includegraphics[width=\columnwidth]{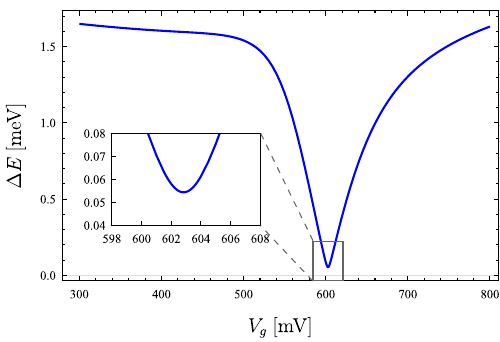}
\caption {\label{fig:orbitals1}Quantum dot orbital splitting as a function of gate voltage. The orbital splitting reaches a minimum value of $54.6$ $\mu$eV. The barrier width is $d=3$ nm and the confinement frequencies are $\hbar\omega_x=2$ meV and $\hbar\omega_y=2.6$ meV.}
\end{figure}

A sufficiently large misalignment of the quantization axes is necessary, but not by itself sufficient, to drive spin rotations via hopping.
 The shuttling process must also be charge-adiabatic to avoid populating excited quantum dot orbital states, yet spin-diabatic to effectively rotate the spin~\cite{loss2026reversible}. %
 Consequently, the inter-well tunnel coupling must significantly exceed the qubit Zeeman splitting, which restricts the applied magnetic field to low values \cite{wang2024operating}.

As a rough estimate for the relevant magnetic-field range and the hopping-gate operation times, we use the criterion
\begin{equation}\label{eq:hopping_crit}
    \frac{\hbar}{\Delta E}
    < t_{\mathrm{shuttle}}
    \lesssim \frac{\hbar}{\Delta E_Z},
\end{equation}
where $\Delta E_Z$ and $\Delta E$ denote the Zeeman splitting and the minimum
orbital splitting, respectively. At the HH--LH resonance,
$\Delta E = 54.6~\mu\mathrm{eV}$, corresponding to
$\hbar/\Delta E = 12.06$ ps. Assuming an effective $g$-factor of $g=2.6$
and an in-plane magnetic field $B_y=10$--$25$ mT, we obtain
$\Delta E_Z=1.50$--$3.76~\mu\mathrm{eV}$, satisfying
$\Delta E_Z \ll \Delta E$. The corresponding Zeeman timescale
$\hbar/\Delta E_Z$ decreases from $437$ ps to $174.9$ ps over this
magnetic-field range. Thus, choosing a shuttling time of approximately
$t_\mathrm{shuttle}=100$ ps is compatible with Eq.~\eqref{eq:hopping_crit}
at the level of this rough estimate. 

Besides the four shuttling steps, a hopping-based gate requires three
free-evolution intervals between the shuttling steps \cite{wang2024operating}.
As a rough estimate, we approximate the total waiting time by three full spin
rotations determined by the Zeeman splitting,
$t_\mathrm{wait}=3h/\Delta E_Z=3.30$--$8.24$ ns.
Together with the four shuttling steps, this gives a total gate time of
$3.70$--$8.64$ ns. We note that the faster gate operation compared with
Ref.~\cite{wang2024operating} is a consequence of the much larger effective
$g$-factor. This, however, requires very fast pulse electronics, which is
experimentally challenging. To relax this requirement, the magnetic field
can be reduced further.

We plot the quantum dot orbital splitting as a function of gate voltage in Fig.~\ref{fig:orbitals1} for a barrier width $d=3~\mathrm{nm}$ and confinement frequencies $\hbar\omega_x=2~\mathrm{meV}$ and $\hbar\omega_y=2.6~\mathrm{meV}$. In the deep LH and HH regimes, the orbital splitting is large, exceeding $1.5~\mathrm{meV}$. However, at the HH-LH resonance, the splitting reaches a minimum value of $54.6~\mu\mathrm{eV}$, which is determined by the inter-well tunnel coupling arising from the thin barrier.

\begin{figure}[htp!]
\centering
\includegraphics[width=\columnwidth]{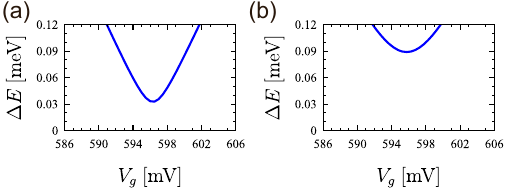}
\caption {\label{fig:orbitals2}Quantum dot orbital splittings as a function of plunger gate voltage, close to the minimum, for a barrier width $d=3$ nm. (a) The confinement frequencies are $\hbar\omega_x=1.5$ meV and $\hbar\omega_y=1.95$ meV. The minimum orbital splitting is 32.7 $\mu$eV. (b) The confinement frequencies are $\hbar\omega_x=1.2$ meV and $\hbar\omega_y=2.4375$ meV. The minimum orbital splitting is 88.8 $\mu$eV.}
\refstepcounter{subfigure}\label{fig:orbitals2a}
\refstepcounter{subfigure}\label{fig:orbitals2b}
\end{figure}
Besides the barrier width, the minimum orbital splitting also depends on the in-plane confinement strength and anisotropy. We find that for isotropic in-plane confinement ($\omega_x=\omega_y$), the orbital splitting vanishes, precluding shuttling
 (as shown by symmetry arguments in Appendix~\ref{App:Confinement}). The stronger and more anisotropic the in-plane confinement potential, the larger the minimum orbital splitting. In Fig.~\ref{fig:orbitals2a}, we plot the orbital splitting for a weaker in-plane confinement compared to Fig.~\ref{fig:orbitals1}, while keeping the ratio $\omega_y/\omega_x$ fixed. For $\hbar\omega_x = 1.5~\mathrm{meV}$, the minimum orbital splitting drops to $32.7~\mu\mathrm{eV}$, compared to the $54.6~\mu\mathrm{eV}$ obtained at $\hbar\omega_x = 2~\mathrm{meV}$. Figure~\ref{fig:orbitals2b} displays the splitting for $\hbar\omega_x = 1.2~\mathrm{meV}$ and $\hbar\omega_y = 2.4375~\mathrm{meV}$. Here, we maintain the same product $\omega_x\omega_y$ as in Fig.~\ref{fig:orbitals2a}, but introduce a larger anisotropy. This configuration yields a large minimum orbital splitting of $88.8~\mu\mathrm{eV}$.

\section{Resonant driving}\label{sec:EDSR}
The conventional way of driving hole spin qubits is based on EDSR, utilizing the intrinsic SOI of holes.
 First, we assume an AC in-plane electric field: 
\begin{equation}
    \mathbf{E}(t)=E_\mathrm{AC}\cos{(\omega t)}[\cos({\varphi}) \hat{x}+\sin{(\varphi)}\hat{y}],
\end{equation}
where $\hat{x}$ ($\hat{y}$) is the unit vector along $x$ ($y$), $E_\mathrm{AC}$ is the amplitude, $\omega$ is the frequency of the oscillating electric field, and $\varphi$ is the in-plane angle of the electric field. Throughout, the direction of the static magnetic field is given by the polar angle $\theta$ measured from the $z$ axis and the azimuthal angle $\phi$ measured from the $x$ axis (see the axes in Fig.~\ref{fig:device}); in-plane fields correspond to $\theta=90\degree$. We consider only the case in which the electric field is parallel to $\hat{x}$ or $\hat{y}$.
 We assume standard EDSR, with $\omega$ resonant with the qubit splitting. The Rabi frequency can be calculated as
\begin{equation}\label{eq:Rabi}
f_\mathrm{Rabi}=\frac{E_\mathrm{AC}\lvert p\rvert}{h}=\frac{eE_\mathrm{AC}\lvert \bra{0}\cos({\varphi})x+\sin({\varphi})y\ket{1}\rvert}{h},
\end{equation}
where $h$ is Planck's constant, $e$ is the elementary charge, $\ket{0}$ and $\ket{1}$ are the qubit states, and $p=e\lvert\bra{0}\cos({\varphi})x+\sin({\varphi})y\ket{1} \rvert$ is the qubit dipole moment.

\begin{figure}[htp!]
\centering
\includegraphics[width=\columnwidth]{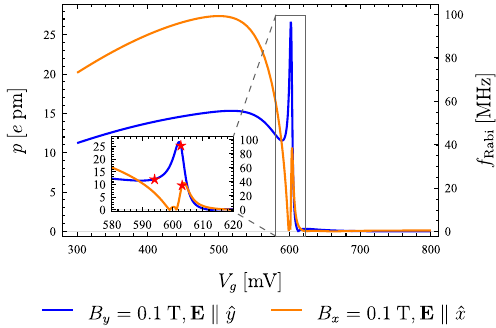}
\caption {\label{fig:Rabi1}In-plane qubit dipole moment and Rabi frequency as a function of plunger gate voltage, with an applied AC electric field along $y$ (blue curve) or $x$ (orange curve), and the in-plane magnetic field $B=0.1$ T parallel to the electric field in both cases. The dipole moment and the Rabi frequency are calculated according to Eq.~\eqref{eq:Rabi}, assuming an electric field amplitude $E_\mathrm{AC}=15$ $\mathrm{mV}/\mu \mathrm{m}$. The barrier width is $d=3$ nm, and the confinement frequencies are $\hbar\omega_x=2$ meV and $\hbar\omega_y=2.6$ meV. The stars show the first-order charge noise sweet spot positions for the respective magnetic field direction.}
\end{figure}

\begin{figure}[htp!]
\centering
\includegraphics[width=\columnwidth]{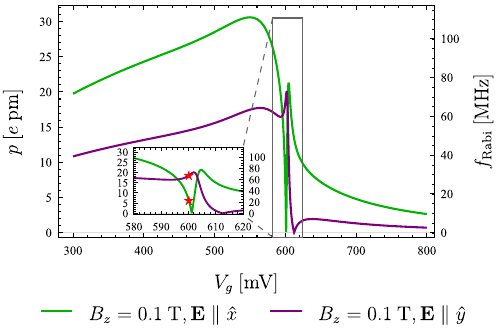}
\caption {\label{fig:Rabi2}Qubit dipole moment and Rabi frequency with an out-of-plane magnetic field as a function of plunger gate voltage, with an AC electric field along $x$ (green curve) or $y$ (purple curve), $B_z=0.1$ T. The dipole moment and the Rabi frequency are calculated according to Eq.~\eqref{eq:Rabi}, assuming an electric field amplitude $E_\mathrm{AC}=15$ $\mathrm{mV}/\mu \mathrm{m}$. The barrier width is $d=3$ nm, and the confinement frequencies are $\hbar\omega_x=2$ meV and $\hbar\omega_y=2.6$ meV. The stars show the first-order charge noise sweet spot positions.}
\end{figure}

We plot the calculated qubit dipole moments and Rabi frequencies in Fig.~\ref{fig:Rabi1} as a function of the plunger gate voltage for in-plane magnetic fields and an in-plane confinement of $\hbar\omega_x=2$~meV and $\hbar\omega_y=2.6$~meV. We assume an electric field amplitude of $E_\mathrm{AC}=15$~mV/$\mu$m, consistent with the setup in Ref.~\cite{froning2021ultrafast}. In the LH regime, the electric field along $x$ yields a larger dipole moment, as this is the weak confinement axis. However, near the HH-LH resonance, the strongly confined $y$ direction yields a peak of the Rabi frequency.
 In the deep HH regime, the linear SOI vanishes, and the dipole moment becomes very small.
Figure~\ref{fig:Rabi2} shows the calculated dipole moments and Rabi frequencies for an out-of-plane magnetic field. When the electric field is parallel to the weak confinement axis, we obtain a larger dipole moment; in both cases, a peak appears near the resonance.
 Note that for both in-plane and out-of-plane magnetic fields, when the electric field is parallel to the weak confinement axis, the dipole moment has a large dip before the peak. The red stars show the positions of first-order charge noise sweet spots. 

We remark that our model neglects shear strains induced by the top gates. Because these strains are known to generate a linear-in-$k$ SOI that enhances the heavy-hole Rabi frequency \cite{abadillo2023hole}, we expect them to provide a further enhancement in the LH regime as well, even though light holes already possess an intrinsic linear-in-$k$ SOI. Furthermore, as shown in Fig.~\ref{fig:Rabi2}, the LH dipole moments obtained in our simulations are much smaller than those predicted in Ref.~\cite{del2023light}. This reduction is a direct consequence of the stronger in-plane confinement assumed in this work, which is necessary to create a large minimum quantum dot orbital level splitting at the HH-LH resonance.

Up to this point, our analysis has focused on in-plane AC electric fields, which laterally displace the wavefunction and drive transitions via the iso-Zeeman EDSR mechanism \cite{golovach2006electric,crippa2018electrical}. In contrast, a purely perpendicular electric field primarily modulates the $g$-tensor, resulting in $g$-tensor modulation resonance ($g$-TMR)~\cite{crippa2018electrical}. In standard single-well heterostructures, the strong vertical confinement suppresses out-of-plane displacement, making this perpendicular driving significantly less efficient than in-plane EDSR. However, this limitation is overcome in our bilayer structure. When the gate voltages are tuned such that the wavefunction delocalizes between the two wells, the system exhibits a large out-of-plane electric dipole moment.
 Consequently, this regime yields large Rabi frequencies, operating in a way analogous to a flopping-mode spin qubit at zero detuning \cite{benito2019electric, croot2020flopping}. This efficient perpendicular driving has also been predicted for electrons in a Si bilayer heterostructure \cite{sarkar2026micromagnet}. 

\begin{figure}[htp!]
\centering
\includegraphics[width=\columnwidth]{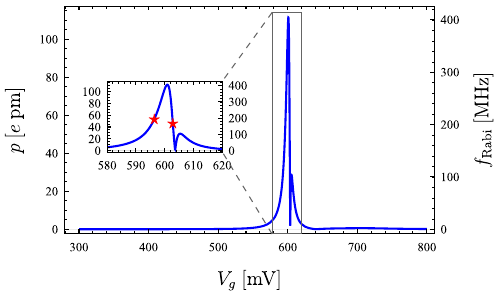}
\caption {\label{fig:Rabi3}Qubit dipole moment and Rabi frequency with an in-plane magnetic field $B=0.1$ T at $\phi=45\degree$ as a function of plunger gate voltage, with an AC electric field along the $z$ direction. We assume an electric field amplitude $E_\mathrm{AC}=15$ $\mathrm{mV}/\mu \mathrm{m}$. The barrier width is $d=3$ nm, and the confinement frequencies are $\hbar\omega_x=2$ meV and $\hbar\omega_y=2.6$ meV.}
\end{figure}

Figure~\ref{fig:Rabi3} shows the calculated out-of-plane dipole moments and Rabi frequencies as a function of gate voltage, for an in-plane magnetic field with $\phi=45\degree$. In the delocalized regime, the Rabi frequencies exceed those for in-plane driving (see Figs.~\ref{fig:Rabi1} and \ref{fig:Rabi2}).
 When the wavefunction is primarily located in one of the wells, the dipole moment becomes negligible. The two stars in Fig.~\ref{fig:Rabi3} show first-order longitudinal charge noise sweet spots for this field orientation, meaning that maximal Rabi frequency and charge noise insensitivity cannot be simultaneously achieved. However, at the charge noise sweet spots, the Rabi frequencies are still large. We also note that for $\mathbf{B}$ at $\phi=45^\circ$, the sweet-spot positions (596.4 and 602.8 mV) differ slightly from those at $\phi=90^\circ$ (594.1 and 602.7 mV). Moreover, they exhibit transverse fluctuations and therefore constitute only longitudinal sweet spots.

\section{Charge noise}\label{sec:noise}
The primary strength of the switchable HH-LH qubit lies in combining the distinct advantages of both regimes. The LH regime, with its large in-plane $g$-factors and fast resonant driving, is ideal for active qubit manipulation. Conversely, in this section, we demonstrate that the HH regime offers superior coherence, making it optimal for idling. Furthermore, we identify a charge noise sweet spot in the highly mixed HH-LH regime, which simultaneously maintains a coherence time comparable to the pure HH regime while still enabling fast Rabi oscillations.

We focus on charge noise, typically the dominant decoherence mechanism in spin qubits \cite{yoneda2018quantum,hendrickx2024sweet}. Instead of simulating individual TLFs, we approximate their collective behavior as an effective plunger gate voltage fluctuation \cite{ReciprocalSweetness}. Assuming that the physically relevant TLFs reside close to the plunger gate, their electrostatic impact is well-captured by this global voltage fluctuation model \cite{piot2022single}. %
 We describe the effect of a plunger gate voltage fluctuation $\delta V(t)$ around $V_0$ via the $g$-tensor: 
\begin{equation}\label{eq:Hqubit}
    H(V_0+\delta V(t))=\frac{1}{2}\mu_B \bm{\sigma} \cdot g(V_0+\delta V(t)) \cdot \bm{B},
\end{equation}
where $\mu_B$ is the Bohr magneton and $\bm{\sigma}$ the vector of Pauli matrices acting on the qubit states. To be able to distinguish the longitudinal and transverse fluctuations of the qubit Hamiltonian from Eq.~\eqref{eq:Hqubit}, we first diagonalize the static qubit Hamiltonian $H(V_0)$ with a unitary transformation $U(V_0)$: 
\begin{equation}
    U(V_0)H(V_0)U^\dag (V_0)=\frac{1}{2}\hbar\omega_0 \sigma_z.
\end{equation}
Using the same unitary matrix $U(V_0)$, we transform the fluctuating Hamiltonian 
\begin{equation}\label{eq:fluctuation}
    \begin{aligned}
        &U(V_0)H(V_0+\delta V(t))U^\dag (V_0)\approx\frac{1}{2}\hbar\omega_0 \sigma_z+\frac{\hbar}{2}D_z \delta V(t)\sigma_z\\
        &+\frac{\hbar}{4} D_{z,2}\delta V(t)^2 \sigma_z+ \frac{\hbar}{2}D_x \delta V(t) \sigma_x+ \frac{\hbar}{2}D_y\delta V(t)\sigma_y,
    \end{aligned}
\end{equation}
where we kept only the longitudinal terms up to second order in $\delta V$ and the first-order transverse terms. Typically, the first-order longitudinal fluctuation $D_z \delta V(t)\sigma_z/2$ dominates dephasing, allowing the remaining terms in Eq.~\eqref{eq:fluctuation} to be neglected \cite{ithier2005decoherence}. Consequently, the dephasing rate is proportional to the derivative of the effective $g$-factor with respect to the plunger gate voltage $V_g$. However, close to the HH-LH resonance, this derivative can vanish (see the red circles in Fig.~\ref{fig:g1}), indicating a first-order charge noise sweet spot. Because of this vanishing first-order susceptibility, a purely first-order approximation is insufficient, and a higher-order description of dephasing becomes necessary to accurately capture the decoherence.

In general, the second-order longitudinal and first-order transverse fluctuations in Eq.~\eqref{eq:fluctuation} are expected to yield comparable decoherence effects. Determining the dominant contribution requires a comparison between the transverse components ($D_x^2/\omega_0$ and $D_y^2/\omega_0$) and the second-order longitudinal term ($D_{z,2}$). We find that when the external magnetic field is aligned with any of the principal confinement symmetry axes ($x$, $y$, or $z$), the transverse fluctuations vanish. However, for an arbitrary magnetic field orientation, these transverse fluctuations become significant.
We assume the magnetic field to be along $y$; therefore, we keep only the first- and second-order longitudinal fluctuations. The phase error caused by the fluctuation $\delta V(t)$ can be written as 
\begin{equation}
    \Delta \phi(t)=D_z \int\displaylimits_0^t \mathrm{d}\tau \delta V(\tau)+\frac{D_{z,2}}{2}\int\displaylimits_0^t \mathrm{d}\tau \delta V^2(\tau).
\end{equation}
The coherence $f_{\Delta \phi}$ is the ensemble average of the accumulated phase factors

\begin{equation}
    f_{\Delta \phi}=\langle e^{i \Delta \phi}\rangle.
\end{equation}
We treat the noise within the quasistatic approximation, assuming that $\delta V$ is static during any single coherent evolution but fluctuates across the statistical ensemble of measurements
\begin{equation}
    \Delta \phi(t)\approx D_z \delta V t+\frac{D_{z,2}}{2}\delta V^2 t.
\end{equation}
We assume a large number of TLFs, resulting in a Gaussian distribution for the fluctuations $\delta V$, with zero mean and standard deviation $\sigma(t)$. The coherence factor can be written as 
\begin{equation}\label{eq:coherence}
\begin{aligned}
    f_{\Delta \phi}(t)&=\frac{1}{\sqrt{2\pi \sigma^2}}\int\displaylimits_{-\infty}^\infty \mathrm{d}\delta V \mathrm{e}^{-\frac{\delta V^2}{2 \sigma^2}}\mathrm{e}^{i\left(D_z \delta V t+\frac{D_{z,2}}{2}\delta V^2 t\right)}\\
    &=\frac{1}{\sqrt{1 - i D_{z,2} \sigma^2 t}} \exp\left[ -\frac{D_z^2 \sigma^2 t^2}{2(1 - i D_{z,2} \sigma^2 t)} \right].
\end{aligned}
\end{equation}
We further assume $1/f$ noise with an infrared cutoff $\omega_\mathrm{ir}$, an ultraviolet cutoff $\omega_\mathrm{uv}$, and power spectral density (PSD)
\begin{equation}
    S(\omega)=\frac{A}{\lvert\omega\rvert}, \hspace{3mm} \omega_\mathrm{ir}<\lvert \omega \rvert <\omega_\mathrm{uv},
\end{equation}
where $A$ is the amplitude of the noise. The PSD is the Fourier transform of the autocorrelation function 
\begin{equation}
    S(\omega)=\int\limits_{-\infty}^{\infty} \mathrm{d}\tau \langle\delta V(t) \delta V(t+\tau)\rangle e^{-i \omega \tau}.
\end{equation}
Since the variance $\sigma^2$ is the autocorrelation function evaluated at $\tau=0$, we obtain

\begin{equation}
    \sigma^2(t)=\frac{1}{2 \pi}\int\displaylimits_{-\infty}^\infty \mathrm{d}\omega S(\omega)=\frac{A}{\pi}\int\displaylimits_{\omega_\mathrm{ir}}\displaylimits^{\omega_\mathrm{uv}} \mathrm{d}\omega\frac{1}{\omega}.
\end{equation}
Under the quasistatic approximation, for short times $t\ll 1/\omega_\mathrm{ir}$, the relevant low-frequency contributions extend up to $\omega \approx 1/t$ \cite{ithier2005decoherence}
\begin{equation}\label{eq:sigma}
    \sigma^2(t)\approx\frac{A}{\pi}\int\displaylimits_{\omega_\mathrm{ir}}^{1/t}\mathrm{d}\omega\frac{1}{\omega}=\frac{A}{\pi}\log{\left(\frac{1}{\omega_\mathrm{ir}t}\right)}.
\end{equation}
By substituting Eq.~\eqref{eq:sigma} into Eq.~\eqref{eq:coherence}, we obtain the complex coherence $f_{\Delta \phi}(t)$. While the absence of the second-order coupling $D_{z,2}$ yields the usual Gaussian decay, operating exactly at the sweet spot ($D_z=0$) results in purely algebraic decay, making the $T_2^*$ coherence time ill-defined. The crossover between these two regimes is continuous; deviations from Gaussian decay become pronounced when the denominator of Eq.~\eqref{eq:coherence} satisfies $D_{z,2}\sigma(t)^2 t = \mathcal{O}(1)$. We consider the decay approximately Gaussian if the following condition is fulfilled

\begin{equation}\label{eq:transition}
    \lvert D_{z,2}\rvert\sigma^2\left(T_2^{*(0)}\right)T_2^{*(0)}<1,
\end{equation}
where $T_2^{*(0)}$ denotes the dephasing time evaluated strictly in the linear limit: first, we consider only the first-order fluctuations and neglect $D_{z,2}$, and then calculate $T_2^{*(0)}$ using the value $D_z$. Subsequently, we take into account the finite $D_{z,2}$ and evaluate the condition in Eq.~\eqref{eq:transition}.

If Eq.~\eqref{eq:transition} is satisfied, we consider the decay to be in the Gaussian regime, and calculate
 $T_2^*$ (not to be confused with $T_2^{*(0)}$) according to
\begin{equation}\label{eq:T2star}
    \lvert f_{\Delta \phi} (T_2^*)\rvert=\frac{1}{\mathrm{e}}. 
\end{equation}
We note that numerically locating the gate voltage $V_g$ at which the crossover in Eq.~\eqref{eq:transition} occurs is challenging.
 First, we assume that in the vicinity of the crossover, $D_{z,2}$ is constant, so $V_g$ is determined by $D_z$ alone. We find the approximate gate voltage values where $D_z$ produces the desired crossover by interpolating the numerically calculated $D_z$.

\begin{figure}[tbh]
\centering
\includegraphics[width=\columnwidth]{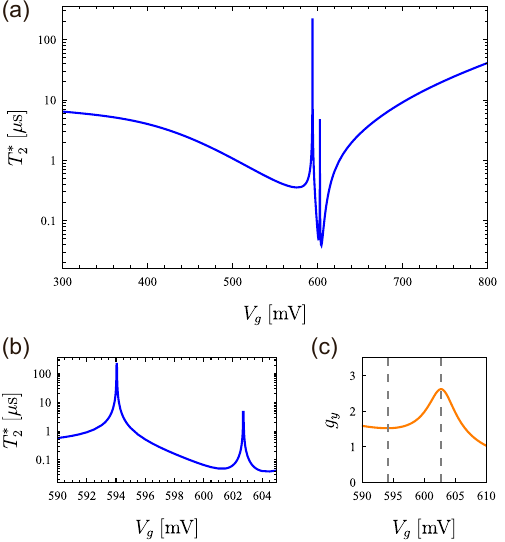}
\caption {\label{fig:noise}Calculated dephasing time as a function of plunger gate voltage. The barrier width is $d=3$ nm, the confinement frequencies are $\hbar\omega_x=2$ meV and $\hbar\omega_y=2.6$ meV. (a) The dephasing time is calculated using Eqs.~\eqref{eq:coherence}, ~\eqref{eq:sigma} and ~\eqref{eq:T2star}, assuming an infrared cutoff frequency $\omega_\mathrm{ir}/2\pi=10^{-2}$ Hz and $\sigma(t=1 \hspace{1mm}\mu\mathrm{s})=10$ $\mu$V. The magnetic field is along $y$, $B_y=0.1$ T. Two peaks appear close to the HH-LH resonance.
 (b) Detailed view of the two dephasing time peaks shown in (a). (c) Corresponding effective $g$-factor as a function of gate voltage. The dashed lines show the positions of the dephasing time peaks shown in (a) and (b).}
\refstepcounter{subfigure}\label{fig:Noisea}
\refstepcounter{subfigure}\label{fig:Noiseb}
\refstepcounter{subfigure}\label{fig:Noisec}
\end{figure}

Figure \ref{fig:Noisea} displays the calculated dephasing time as a function of gate voltage, according to the condition in Eq.~\eqref{eq:transition}, assuming an infrared cutoff frequency $\omega_{\mathrm{ir}}/2\pi = 10^{-2}$ Hz, and gate voltage fluctuations $\sigma(t=1\hspace{1mm}\mu\mathrm{s})=10$ $\mu$V. As expected, the coherence time in the deep HH regime is approximately six times longer than in the LH regime. Notably, two distinct peaks emerge near the HH-LH resonance, shown in detail in Fig.~\ref{fig:Noiseb}. As illustrated in Fig.~\ref{fig:Noisec}, these dephasing time peaks coincide with the zeros of the first derivatives of the effective $g$-factors (i.e., first-order sweet spots). Crucially, our second-order noise calculation reveals that the sweet spot at the lower gate voltage yields a longer dephasing time due to its smaller $g$-factor curvature. Consequently, for optimal qubit operation, this lower-voltage sweet spot—corresponding to approximately 25\% HH weight in the wavefunction—is preferable to the 50\% HH resonance point at $V_g = 602.7$ mV.

In the immediate vicinity of the two charge-noise sweet spots, the condition given by Eq.~\eqref{eq:transition} breaks down. Consequently, Fig.~\ref{fig:noise} includes only the data points that satisfy this criterion. Under this restriction, we extract a dephasing time of $T_2^*=217.6$~$\mu$s ($4.7$~$\mu$s) for the left (right) peak. We emphasize that naively applying Eq.~\eqref{eq:T2star} exactly at the left peak ($D_z=0$) yields a purely algebraic decay, for which $T_2^*$ is ill-defined; formally, Eq.~\eqref{eq:T2star} then gives a value exceeding 1~ms.
 Furthermore, the remarkably long coherence time of $217.6$~$\mu$s predicted here is a consequence of neglecting hyperfine interactions, which may become the limiting mechanism at the sweet spot. This limitation is not fundamental, however: hyperfine noise can be strongly suppressed by isotopic purification, since all three constituent elements possess abundant spin-zero isotopes---the spin-carrying isotopes $^{29}$Si, $^{73}$Ge, and $^{115,117,119}$Sn make up only 4.7\%, 7.8\%, and 16.6\% of the natural elements, respectively~\cite{itoh2014isotope}. Moreover, since the hole resides predominantly in the Ge quantum wells, purifying the Ge layers alone already removes most of the hyperfine coupling. %

Furthermore, Fig.~\ref{fig:Noiseb} reveals regions of strongly suppressed dephasing time immediately adjacent to the sweet spots, corresponding to the steep $g$-factor gradients shown in Fig.~\ref{fig:Noisec}. Traversing this highly noise-sensitive region to switch between the LH and HH states demands rapid gate voltage sweeps to avoid decoherence. Nevertheless, the impact of quasistatic noise during this operation may be naturally mitigated: because the $g$-factor derivative inverts its sign across the resonance peak at $V_g=602.7$ mV, the phase error accumulated on the left side of the transition is expected to partially cancel the error accumulated on the right.

Finally, we emphasize that our noise model is defined by two effective parameters: the infrared cutoff frequency $\omega_{\mathrm{ir}}$ and the noise amplitude $A$. Because the specific value of $A$ is highly sensitive to the quality and purity of the experimental heterostructure, the exact magnitude of the dephasing time may vary. Therefore, rather than providing absolute predictions for coherence, Fig.~\ref{fig:noise} is primarily intended to illustrate the relative dephasing times across the different operational regimes.

\section{Discussion}\label{sec:discussion}
In this work, we proposed a bilayer Ge heterostructure with SiGeSn barriers to implement a switchable HH-LH qubit, combining the high operation speeds of light holes with the superior coherence of heavy holes. This platform highlights the broad potential of SiGeSn for valence-band and strain engineering, enabling the design of spin-qubit architectures with new functionalities. The proposed device relies on engineering a weak tensile strain to induce a small HH-LH splitting and establish an LH ground state. Crucially, strong quantum-well confinement can overcome this splitting, resulting in an HH ground state. Creating this system requires a careful balance in the SiGeSn composition: the Si concentration must be high enough to provide large valence-band offsets for confinement, while maintaining a lattice constant slightly larger than that of Ge to induce the requisite tensile strain. While experimentally demanding, we estimate that achieving the target strain requires controlling the Si concentration to a precision of approximately ±1\%, which is an accessible target for modern epitaxial growth techniques \cite{zhang2026numerical}. The allowed uncertainty reflects the requirement that the strain induces a sufficiently small HH--LH splitting for the confinement in the thin Ge well to overcome it. At the same time, in the deep HH (LH) regime, corresponding to $V_g=800$ mV ($V_g=300$ mV), the HH--LH splitting must remain sufficiently large; otherwise, the HH (LH) subband develops deep minima at finite $k$ rather than at the $\Gamma$ point. We find that these two requirements constrain the in-plane strain to approximately $\varepsilon_{xx}=0.20\%$--$0.29\%$. 

Tuning the vertical electric field via a plunger gate allows switching of the qubit between LH and HH regimes, facilitating single-qubit operations through hopping or resonant driving. In contrast to the concurrent proposal of Ref.~\cite{valvo2026lightholeARXIV}, where the LH state resides in a strained SiGe well, both quantum wells in our design consist of pure Ge, keeping alloy disorder out of the channel at the cost of a ternary barrier. At the transition, in-plane confinement drives an HH-LH resonance characterized by a strongly mixed HH-LH state, pronounced in-plane $g$-factor peaks, and a pair of first-order charge noise sweet spots. A second-order noise analysis identifies the sweet spot with the lower HH fraction as the optimal operating point, owing to its smaller $g$-factor curvature. This point maintains a Rabi frequency similar to the deep LH regime while extending the dephasing time by an order of magnitude. Coherent shuttling across this transition, however, demands rapid gate-voltage sweeps to minimize decoherence in the highly noise-sensitive regions surrounding the sweet spots.

In conclusion, this study introduces SiGeSn as an unexplored material platform for hole spin qubits, enabling novel devices such as the switchable HH-LH spin qubit. As the first work to propose SiGeSn for this application, we highlight its unique capacity for heterostructure engineering. However, more accurate theoretical modeling of these nanostructures is currently constrained by the absence of established Luttinger parameters for SiGeSn alloys. Determining these band-structure parameters will be an important next step toward refining the modeling of future SiGeSn-based quantum devices.

\begin{acknowledgments}
We thank Zoé McIntyre, Abhikbrata Sarkar, Patrick del Vecchio, and Stefano Bosco for useful discussions. 
This work was supported as part of NCCR SPIN, a National Center of Competence in Research, funded by the Swiss National Science Foundation (grant number 225153). This work has received funding from the Swiss State Secretariat for Education, Research and Innovation (SERI) under contract number M822.00078.  The Gen-Q programme has received funding from the European Union’s Horizon Europe research and innovation programme under the Marie Skłodowska-Curie grant agreement number 101217386. Calculations were performed at sciCORE (http://scicore.unibas.ch/) scientific computing center at the University of Basel. D.L. acknowledges the Deanship of Research and the Quantum Center at KFUPM for the support received under Grant no. CUP25102 and no. INQC2600, respectively.
\end{acknowledgments}

\appendix

\section{LKBP Hamiltonian and material parameters}\label{App:parameters}
We model the heterostructure presented in Fig.~\ref{fig:device} using the 6-band LKBP Hamiltonian \cite{luttinger1956quantum,bir1974symmetry,winkler2001spin}
\begin{equation}
    H_\mathrm{LKBP}=H_\mathrm{LK}+H_\mathrm{BP},
\end{equation}
where $H_\mathrm{LK}$ is the Luttinger-Kohn Hamiltonian containing the kinetic terms 
\begin{widetext}
\begin{equation}\label{eq:HLK}
  H_\mathrm{LK}=\begin{pmatrix}
      P+Q & -S & R & 0 & S/\sqrt{2} & -\sqrt{2}R \\ 
      -S^* & P-Q & 0 & R & \sqrt{2}Q & -\sqrt{3/2} S\\ 
      R^* & 0 & P-Q & S & -\sqrt{3/2} S^* & -\sqrt{2}Q \\
      0 & R^* & S^* & P+Q & \sqrt{2}R^* & S^*/\sqrt{2} \\ 
      S^*/\sqrt{2} & \sqrt{2} Q^* & -\sqrt{3/2}S & \sqrt{2}R & P & 0 \\ 
      -\sqrt{2}R^* & -\sqrt{3/2}S^* & -\sqrt{2}Q^* & S/\sqrt{2} & 0 & P
  \end{pmatrix},
\end{equation}
with elements 
\begin{equation}
\begin{aligned}
    &P=\frac{\hbar^2}{2m_0}\gamma_1(k_x^2+k_y^2+k_z^2), \hspace{5mm} 
    Q=-\frac{\hbar^2}{2m_0}\gamma_2(2k_z^2-k_x^2-k_y^2), \\
    &R=\sqrt{3}\frac{\hbar^2}{2m_0}[-\gamma_2(k_x^2-k_y^2)+2i\gamma_3 k_xk_y], \hspace{5mm} S=\sqrt{3}\frac{\hbar^2}{m_0}\gamma_3(k_x-ik_y)k_z,  
\end{aligned}
\end{equation}
where $\gamma_1$, $\gamma_2$, $\gamma_3$ are the Luttinger mass parameters. Assuming biaxial strain, the Bir-Pikus Hamiltonian has the form 
\begin{equation}\label{eq:HBPmatrix}
    H_\mathrm{BP}=-a_v\mathrm{Tr\epsilon}-b_v(\epsilon_{xx}-\epsilon_{zz})\begin{pmatrix}
        1 & 0 & 0 & 0 & 0 & 0 \\ 
        0 & -1 & 0 & 0 & \sqrt{2} & 0 \\ 
        0 & 0 & -1 & 0 & 0 & -\sqrt{2} \\
        0 & 0 & 0 & 1 &  0& 0 \\
        0 & \sqrt{2} & 0 & 0 & 0 & 0 \\
        0 & 0 & -\sqrt{2} &0 & 0 & 0
    \end{pmatrix},
\end{equation}
with $\epsilon_{xy}=\epsilon_{xz}=\epsilon_{yz}=0$, $\epsilon_{xx}=\epsilon_{yy}$, $\epsilon_{zz}=-2 \epsilon_{xx} C_{12}/C_{11}$, with elastic constants $C_{11}$ and $C_{12}$ and deformation potentials $a_v$ and $b_v$. The Hamiltonians from Eqs.~\eqref{eq:HLK} and~\eqref{eq:HBPmatrix} are written in the basis of heavy holes, light holes and split-off holes, $\ket{\frac{3}{2},\frac{3}{2}}$, $\ket{\frac{3}{2},\frac{1}{2}}$, $\ket{\frac{3}{2},-\frac{1}{2}}$, $\ket{\frac{3}{2},-\frac{3}{2}}$, $\ket{\frac{1}{2},\frac{1}{2}}$ and $\ket{\frac{1}{2},-\frac{1}{2}}$. 
\end{widetext}
We also add the direct Zeeman Hamiltonian 
\begin{equation}
    H_Z=\begin{pmatrix}
        H_Z^\mathrm{(HL)} & H_Z^{(C)} \\[2mm] 
        H_Z^{(C)\dag} & H_Z^\mathrm{(SO)}
    \end{pmatrix}, 
\end{equation}
where $H_Z^\mathrm{(HL)}$ is the Zeeman term of the heavy and light holes 
\begin{equation}
    H_Z^\mathrm{(HL)}=2\mu_B (\kappa \bm{J}\cdot \bm{B}+q\bm{\mathcal{J}}\cdot \bm{B}), 
\end{equation}
where $\kappa$ and $q$ are material-dependent Zeeman constants, $\bm{J}=(J_x,J_y,J_z)$, and $\bm{\mathcal{J}}=(J_x^3,J_y^3,J_z^3)$, and $J_i$ ($i\in \{x,y,z\}$) are the 3/2 spin matrices. The Zeeman term for split-off holes is 
\begin{equation}
    H_Z^\mathrm{(SO)}=
        2\mu_B \kappa \bm{\sigma}\cdot \bm{B}, 
\end{equation}
while the coupling
\begin{equation}
    H_Z^{(C)}=3 \mu_B \kappa\bm{U}\cdot \bm{B}, 
\end{equation}
where $\bm{U}=(U_x,U_y,U_z)$ with components: 
\begin{subequations}
\begin{align}
    U_x&=\frac{1}{3\sqrt{2}}\begin{pmatrix}
        -\sqrt{3} & 0 \\ 
        0 & -1\\
        1 & 0\\
        0 & \sqrt{3}
    \end{pmatrix},\\[2mm]
    U_y&=\frac{i}{3\sqrt{2}}\begin{pmatrix}
        \sqrt{3} & 0 \\
        0 & 1 \\ 
        1 & 0 \\
        0 & \sqrt{3}
    \end{pmatrix},\\[2mm]
    U_z&=\frac{\sqrt{2}}{3} \begin{pmatrix}
        0 & 0 \\ 
        1 & 0 \\ 
        0 & 1 \\ 
        0 & 0
    \end{pmatrix}.
\end{align}
\end{subequations}
The orbital effects are taken into account via the substitution
\begin{equation}
    \bm{k}=-i\nabla+\frac{e}{\hbar}\bm{A},
\end{equation}
where $\bm{A}$ is the vector potential
\begin{equation}
    \bm{A}=(z B_y-y B_z,-zB_x,0).
\end{equation}
The quantum well confinement Hamiltonian $V_\mathrm{qw}(z)$ has the form \cite{del2023light,del2024light,del2025fully}
\begin{equation}\label{eq:QW1}
    V_\mathrm{qw}=-E_V-V_\mathrm{SO}, 
\end{equation}
where $E_V$ is the average valence-band energy, and $V_\mathrm{SO}$ is given by
 
\begin{equation}\label{eq:VSO}
V_\mathrm{SO}=\frac{\Delta_\mathrm{SO}}{3}\mathrm{Diag}\left(1,1,1,1,-2,-2\right),
\end{equation}
which provides the split-off gap $\Delta_\mathrm{SO}$ between the split-off holes and the heavy and light holes.

We assume unstrained SiGeSn barriers, while the strain in the Ge quantum wells can be calculated as 
\begin{equation}
    \epsilon_{xx,\mathrm{Ge}}=\frac{a_\mathrm{SiGeSn}}{a_\mathrm{Ge}}-1, 
\end{equation}
where $a_\mathrm{Ge}$ is the lattice constant of Ge, while $a_\mathrm{SiGeSn}$ is the lattice constant of $\mathrm{Si}_x\mathrm{Ge}_{1-x-y}\mathrm{Sn}_y$ calculated as
 
\begin{equation}
\begin{aligned}
    a_\mathrm{SiGeSn}=&x a_\mathrm{Si}+y a_\mathrm{Sn}+(1-x-y) a_\mathrm{Ge}\\&-b_\mathrm{SiGe}x(1-x-y)-b_\mathrm{SnGe}y(1-x-y),
\end{aligned}
\end{equation}
where $a_\mathrm{Si(Sn)}$ is the lattice constant of Si (Sn), $b_\mathrm{SiGe}$ and $b_\mathrm{SnGe}$ are bowing parameters, $b_\mathrm{SiGe}=0.0188$ \AA{} \cite{fischetti1996band}, $b_\mathrm{SnGe}=-0.083$ \AA \cite{polak2017electronic}. We omit the bowing parameter of SiSn, as it is found to be negligible \cite{tolle2006low, moontragoon2012direct}.
 The remaining material parameters are linearly interpolated between the values of Si, Ge, and Sn, which can be found in Table~\ref{tab:parameters}. 

\begin{table}[htp!]
\centering
\caption{Material parameters used in the calculations. The SiGeSn values are linearly interpolated between the pure material parameters.}\label{tab:parameters}
\renewcommand{\arraystretch}{1.3} %
\begin{tabular}{lccc}
\toprule
 & \textbf{Ge} & \textbf{Si} & \textbf{Sn} \\ 
\midrule

\multicolumn{4}{l}{Lattice constant} \\
$a_0$ \quad [\AA] & $5.65235^{\mathrm{a}}$ & $5.42982^{\mathrm{a}}$ & $6.480117^{\mathrm{b}}$ \\
\hline

\multicolumn{4}{l}{Bulk band energies} \\
$E_V$ [eV] & 0 & $-0.49^{\mathrm{j}*}$ & $0.69^{\mathrm{h}*}$ \\
$\Delta_\mathrm{SO}$ \quad [eV] & $0.290^{\mathrm{b}}$ & $0.044^{\mathrm{i}}$ & $0.770^{\mathrm{f}}$ \\
\hline

\multicolumn{4}{l}{Elastic constants} \\
$C_{12}/C_{11}$ & $0.333^{\mathrm{b}}$ & - & - \\
\hline

\multicolumn{4}{l}{Deformation potentials} \\
$a_v$ \quad [eV] & $1.24^{\mathrm{c}}$ & - & - \\
$b_v$ \quad [eV] & $-2.86^{\mathrm{d}}$ & - & - \\
\hline

\multicolumn{4}{l}{Effective mass and spin parameters} \\
$\gamma_1$ & $13.38^{\mathrm{e}}$ & $\dagger$ & $\dagger$ \\
$\gamma_2$ & $4.24^{\mathrm{e}}$ & $\dagger$ & $\dagger$ \\
$\gamma_3$ & $5.69^{\mathrm{e}}$ & $\dagger$ & $\dagger$ \\
$\kappa$ & $3.41^{\mathrm{f}}$ & $\dagger$ & $\dagger$ \\
$q$ & $0.06^{\mathrm{e}}$ & $\dagger$ & $\dagger$ \\
\bottomrule

\multicolumn{4}{l}{
  \begin{tabular}[t]{@{}l@{}}
  References: $^{\mathrm{a}}$:\cite{reeber1996thermal}, $^{\mathrm{b}}$:\cite{raasander2015accuracy}, $^{\mathrm{c}}$:\cite{van1989band}, \\
  $^{\mathrm{d}}$:\cite{van1986theoretical}, $^{\mathrm{e}}$:\cite{winkler2001spin}, $^{\mathrm{f}}$:\cite{lawaetz1971valence}, $^{\mathrm{h}}$:\cite{menendez2004type}, \\
  $^{\mathrm{i}}$:\cite{fischetti1996band}, $^{\mathrm{j}}$:\cite{edward1990measurement} \\
  $^*$ Relative to Ge \\
  $^\dagger$ Ge values used\\ 
  \end{tabular}
}
\end{tabular}
\par\smallskip
\begin{minipage}{\columnwidth}\raggedright

\end{minipage}
\end{table}

\section{Alternative heterostructure layer ordering}\label{App:Alternative}

\begin{figure}[htp!]
\centering
\includegraphics[width=\columnwidth]{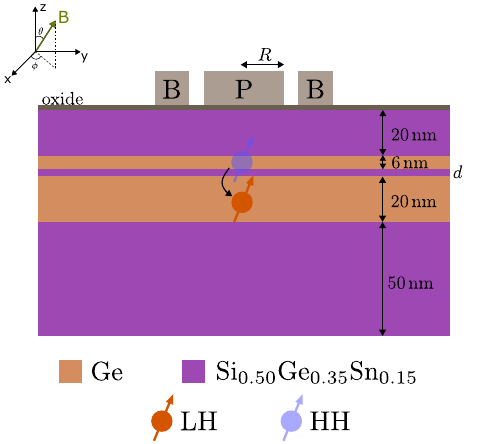}

\caption {\label{fig:device2}Alternative heterostructure layer ordering compared to Fig.~\ref{fig:device}.
 In this layout, the HH well is positioned closer to the top gates. For better electrostatic confinement, the upper barrier layer has been reduced from 30 nm to 20 nm.}
\end{figure}

A limitation of the initial heterostructure (Fig.~\ref{fig:device}) is that it necessitates positive plunger and negative barrier gate voltages. We therefore present an alternative layout featuring an inverted layer order, placing the HH quantum well above the LH well. As illustrated in Fig.~\ref{fig:device2}, alongside swapping the quantum wells, we also reduce the thickness of the upper barrier from 30 nm to 20 nm. This additional modification is necessary for good confinement by the plunger gate potential in Eq.~\eqref{eq:Vz}. 

\begin{figure}[htp!]
\centering
\includegraphics[width=\columnwidth]{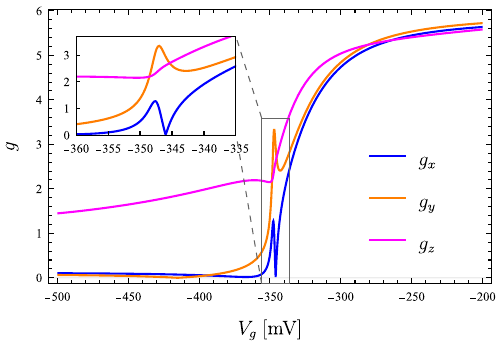}

\caption {\label{fig:NewDesigng}Effective $g$-factors as a function of plunger gate voltage for the alternative heterostructure shown in Fig.~\ref{fig:device2}. The $g$-factor peaks are similar to the peaks observed for the original heterostructure, see Fig.~\ref{fig:g1}. The barrier width is $d=3$ nm, the in-plane confinement frequencies are $\hbar\omega_x=2$ meV and $\hbar\omega_y=2.6$ meV.}
\end{figure}

As in Sec.~\ref{sec:g-factor}, we calculate the $g$-tensors corresponding to the heterostructure in Fig.~\ref{fig:device2}. We plot the effective $g$-factors in Fig.~\ref{fig:NewDesigng}. In contrast to Fig.~\ref{fig:g1}, here the left (right) side of the plot corresponds to the HH (LH) regime. Aside from this, the observed $g$-factor peaks are very similar to those of the original heterostructure design.

\begin{figure}[H]
\centering
\includegraphics[width=\columnwidth]{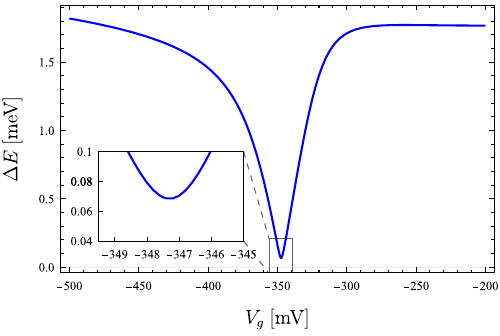}

\caption {\label{fig:NewDesignOrb}Quantum dot orbital level splitting for the alternative heterostructure design shown in Fig.~\ref{fig:device2}. The minimum orbital level splitting is 68.6 $\mu$eV, even larger than for the original heterostructure, see Fig.~\ref{fig:orbitals1}. The barrier width is $d=3$ nm, the in-plane confinement frequencies are $\hbar\omega_x=2$ meV and $\hbar\omega_y=2.6$ meV.}
\end{figure}

In Fig.~\ref{fig:NewDesignOrb}, we show the orbital level splitting of the quantum dot as a function of plunger gate voltage. The minimum orbital level splitting is 68.6 $\mu$eV, larger than the 54.6 $\mu$eV of the original heterostructure (Fig.~\ref{fig:orbitals1}).

\section{Orbital splitting and isotropic in-plane confinement}\label{App:Confinement}
In this Appendix, we argue that an isotropic in-plane confinement leads to a zero minimum quantum dot orbital level splitting, detrimental to shuttling. Let us set the magnetic field to zero. In the case of isotropic confinement, $\omega_x=\omega_y$, the full, 3D Hamiltonian has four-fold in-plane rotation symmetry, which is represented by the operator $R$ that commutes with the Hamiltonian

\begin{equation}
    [H_\mathrm{3D},R]=0.
\end{equation}
This means that $H_\mathrm{3D}$ and $R$ have common eigenstates 
\begin{equation}
    R\ket{\Psi_n}=r_n\ket{\Psi_n}. 
\end{equation}
If we apply the rotation four times to the spinor, we obtain: 
\begin{equation}\label{eq:rn}
    R^4\ket{\Psi_n}=-\ket{\Psi_n} \rightarrow r_n^4=-1, 
\end{equation}
where the minus sign is a consequence of the half-integer spin of the holes.
 Equation~\eqref{eq:rn} yields two sets of $r_n$ values, $r_n=(1\pm i)/\sqrt{2}$ and $r_n=-(1\pm i)/\sqrt{2}$. In the absence of magnetic field, the system also has time-reversal symmetry, represented by the antiunitary operator $\mathcal{T}$, which commutes with rotations
 
\begin{equation}\label{eq:TR}
    [\mathcal{T},R]=0. 
\end{equation}
On a state with $R\ket{\Psi_n}=r_n\ket{\Psi_n}$, $\mathcal{T}$ acts as 
\begin{equation}\label{eq:TR2}
\mathcal{T}R\ket{\Psi_n}=r_n^*\mathcal{T}\ket{\Psi_n}=R \mathcal{T}\ket{\Psi_n},
\end{equation}
where we used Eq.~\eqref{eq:TR} and the antiunitarity of $\mathcal{T}$. Equation~\eqref{eq:TR2} means that for a state $\ket{\Psi_n}$ with eigenvalue $r_n$, its Kramers partner has $r_n^*$. For this analysis, we focus on the four lowest-energy states of the quantum dot, comprising the first two orbital levels and their Kramers-degenerate spin pairs. We consider the following effective two-level Hamiltonian in the basis of two diabatic quantum-dot states, $\ket{\Psi_1}$ and $\ket{\Psi_2}$, predominantly localized in the lower and upper quantum wells, respectively
\begin{equation}\label{eq:Heff_orbitals}
    H_\mathrm{eff}(V_g)=\begin{pmatrix}
        E_1(V_g) & t(V_g) \\
        t^*(V_g) & E_2(V_g)
    \end{pmatrix}, 
\end{equation}
where $E_{1,2}(V_g)$ are the diabatic energies of the two localized states, and $t(V_g)$ is the inter-well tunneling, all depending on the plunger gate voltage. At the gate voltage corresponding to the minimum orbital level splitting, the two diagonal elements are equal, $E_1=E_2$. If there is a finite tunneling $t$, an anticrossing opens. Using the identity $H=R^\dag H R$, the tunneling matrix element can be written as
\begin{equation}\label{eq:tr}
    t=\bra{\Psi_1} H_\mathrm{3D}\ket{\Psi_2}=\bra{\Psi_1}R^\dag H_\mathrm{3D} R \ket{\Psi_2}=r_1^* r_2 t, 
\end{equation}
where we assumed that the effective Hamiltonian in Eq.~\eqref{eq:Heff_orbitals} is the projection of the full Hamiltonian $H_\mathrm{3D}$ on the states $\ket{\Psi_1}$ and $\ket{\Psi_2}$. From Eq.~\eqref{eq:tr} follows that 
\begin{equation}
    t(1-r_1^*r_2)=0. 
\end{equation}
which leads to $t=0$, provided that $r_1\neq r_2$. The lowest HH and LH quantum-dot orbitals carry no envelope orbital angular momentum, while the HH and LH components have different angular-momentum projections. Consequently, $\ket{\Psi_1}$ and $\ket{\Psi_2}$ transform differently under fourfold rotations, implying $r_1\neq r_2$ and hence $t=0$.

For an anisotropic in-plane confinement, $\omega_x\neq \omega_y$, the four-fold rotational symmetry is reduced to two-fold. In this case, $r_n^2=-1$, $r_n=\pm i$. The states $\ket{\Psi_1}$ and $\ket{\Psi_2}$ are not Kramers partners, $r_1=r_2=i$ or $r_1=r_2=-i$, leading to $(1-r_1^*r_2)=0$. In this case, $t\neq 0$ is allowed. 

\bibliography{references} 

\end{document}